\documentclass[english]{article}
\usepackage[T1]{fontenc}
\usepackage[latin9]{inputenc}
\usepackage{amsmath}
\usepackage{babel}
\begin{document}

\title{A Concise Information-Theoretic Derivation of the Baum-Welch algorithm}

\author{Alireza Nejati, Charles Unsworth}
\maketitle
\begin{abstract}
We derive the Baum-Welch algorithm for hidden Markov models (HMMs)
through an information-theoretical approach using cross-entropy instead
of the Lagrange multiplier approach which is universal in machine
learning literature. The proposed approach provides a more concise
derivation of the Baum-Welch method and naturally generalizes to multiple
observations.
\end{abstract}

\subsection*{Introduction}

The basic hidden Markov model (HMM)\cite{rabiner1989tutorial} is
defined as having a sequence of \emph{hidden} or \emph{latent} states
$Q=\{q_{t}\}=\{q_{1},q_{2},\dots,q_{T}\}$ (where $t$ denotes time
interval), and each state is statistically independent of all but
the state immediately before it and where each state emits some observation
$o_{t}$ with a stationary (non-time-varying) probability density.
Formally, the model is defined as:

\begin{equation}
p(O,Q|\lambda)=p(q_{1}|\lambda)\left[\prod_{t=2}^{T}p(q_{t}|q_{t-1},\lambda)\right]\left[\prod_{t=1}^{T}p(o_{t}|q_{t},\lambda)\right]\label{eq:obsdens}
\end{equation}

Where $\lambda=(\pi,a,b)$ is a set of model parameters. The probability
of being in an initial state $q_{1}$ is given by the function $p(q_{1}|\lambda)=\pi(q_{1})$.
The probability of transitioning from state $q_{t}$ to state $q_{t+1}$
is given by $p(q_{t}|q_{t-1},\lambda)=a(q_{t},q_{t-1}).$ Finally,
the emission density is $p(o_{t}|q_{t})=b(o_{t},q_{t})$. Given an
observation sequence $O$, it is desirable to find a set of parameters
that will maximize the likelihood of producing $O$. Generally, finding
the most optimal set of parameters may be computationally difficult;
an approximation is to use the well-known Expectation-Maximization
(EM) algorithm \cite{baum1972}. The special case of the EM algorithm
applied to hidden Markov models is known as the Baum-Welch algorithm
\cite{baum1972}\cite{Fine1998}. The usual approach to deriving the
Baum-Welch is through the use of Lagrange multipliers \cite{Li2000b}\cite{park1998truly}.
In this article, we demonstrate that this approach can be improved
upon using a method based on cross-entropy that is more concise and
lends itself more easily to various HMM generalizations, such as multiple
observation sequences.

\subsection*{The Expectation-Maximization Algorithm}

For the sake of notational simplicity, we will use $E_{Q}[.]$ to
mean the expected value of the expression inside the brackets over
$Q$ given the data and the prior model: ($O,\lambda')$. The EM algorithm
is as follows. Given an existing set of model parameters $\lambda'$
and set of observations $O$, find a new set of model parameters $\lambda$
such that the following function is maximized:

\begin{equation}
\mathcal{Q}(\lambda,\lambda')=E_{Q}[\log p(O,Q|\lambda)]
\end{equation}

By (\ref{eq:obsdens}), we may rewrite this as:

\begin{equation}
\mathcal{Q}(\lambda,\lambda')=E_{Q}\left[\log p(q_{1}|\lambda)\right]+\sum_{t=2}^{T}E_{Q}\left[p(q_{t}|q_{t-1},\lambda)\right]+\sum_{t=1}^{T}E_{Q}\left[\log p(o_{t}|q_{t},\lambda)\right]\label{eq:aux1}
\end{equation}

For each term, we now only use the components of $\lambda=(\pi,a,b)$
that the term depends on and take the expectation over the time-steps
of $q$ that are used:

\begin{multline}
\mathcal{Q}(\lambda,\lambda')=E_{q_{1}}\left[\log p(q_{1}|\pi)\right]+\\
\sum_{t=2}^{T}E_{(q_{t-1},q_{t})}\left[\log p(q_{t}|q_{t-1},a)\right]+\sum_{t=1}^{T}E_{q_{t}}\left[\log p(o_{t}|q_{t},b)\right]\label{eq:aux}
\end{multline}

We note, now, that each term involves optimization of a separate variable
and thus the terms can be optimized separately. For the first term
in (\ref{eq:aux}), $E_{q_{1}}\left[\log p(q_{1}|\pi)\right]=-H(p(q_{1}|O,\lambda'),p(q_{1}|\pi))$,
where H denotes cross entropy. Thus, one must find a $\pi$ that will
minimize the cross entropy between $p(q_{1}|O,\lambda')$ and $p(q_{1}|\pi)$.
To minimize cross entropy, it suffices to set the distributions to
be equal i.e. set $\pi_{i}$ to be $P(q_{1}=i|O,\lambda')$ for all
states $i$. This value can be computed using the forward-backward
algorithm.

The procedure for the second term in (\ref{eq:aux}) is similar. By
writing out the expectation explicitly, the following is obtained:

\begin{multline}
\sum_{t=2}^{T}E_{(q_{t-1},q_{t})}\left[\log p(q_{t}|q_{t-1},a)\right]=\\
\sum_{t=2}^{T}\sum_{i=1}^{N}\sum_{j=1}^{N}p(q_{t}=j,q_{t}=i|O,\lambda')\log p(q_{t}=j|q_{t-1}=i,a_{ij})
\end{multline}

Noting that the term in front of the logarithm is independent of $t$,
we may rewrite this as follows:

\begin{equation}
=\sum_{i=1}^{N}\sum_{j=1}^{N}\log p(q_{2}=j|q_{1}=i,a_{ij})\sum_{t=2}^{T}p(q_{t}=j,q_{t}=i|O,\lambda')
\end{equation}

Now define the new density $\alpha_{i}$ such that $\forall j\: P(q_{t}=j)=\frac{1}{\gamma_{i}}\sum_{t=2}^{T}p(q_{t}=j,q_{t-1}=i|O,\lambda')$
where $\gamma_{i}$ is some normalizing constant to make this a probability
density. Now, we may rewrite the previous equation in terms of expectations:

\begin{equation}
=\sum_{i=1}^{N}E_{\alpha_{i}}[\log p(q_{2}|q_{1}=i,a_{ij})]
\end{equation}

Additionally, define the density $a_{i}$ to be the denstity of $P(q_{2}=j)$
given $a$ and $q_{1}=i$. Thus, the above becomes simply $-\sum_{i=1}^{N}H(\alpha_{i},p(q_{2}|a_{i}))$.
Here we are minimizing a sum of independent cross-entropies (each
term in the sum is independent since the $a_{i}$'s are independent)
which is minimized by setting the density $a_{i}$ to $\alpha_{i}$.
That is, $a_{ij}=\frac{1}{\gamma_{i}}\sum_{t=2}^{T}p(q_{t}=j,q_{t-1}=i|O,\lambda')$,
where $\gamma_{i}=\sum_{j=1}^{N}a_{ij}$. These values may be computed,
again, using a forward and backward iteration through the chain.

The third term in (\ref{eq:aux}) depends on the probability density
used for $b$, and takes on different forms for discrete, gaussian,
or other types of emission distributions. However, in all cases, it
is simply equal to the $\mathcal{Q}$ function in the EM algorithm
for the mixture model using that specific distribution in question:

\begin{equation}
\mathcal{Q}(b,b')=E_{Q}\left[\log\prod_{t=1}^{T}p(o_{t}|q_{t},b)\right]
\end{equation}

Thus, the problem reduces to performing an expectation-maximization
iteration for a mixture model with $N$ mixture components ($N$ =
number of possible states for $q$), where each mixture component
has density $p$, independent of the HMM.

\subsection*{Multiple observations.}

The extension of the above algorithm to multiple independent observation
sequences \cite{Li2000b}, then, is straightforward. Consider the
case of two observation sequences. The $\mathcal{Q}$ function becomes:

\begin{equation}
\mathcal{Q}(\lambda,\lambda')=E_{q^{(1)}q^{(2)}|(O^{(1)},O^{(2)},\lambda')}[\log p(O^{(1)},q^{(1)},O^{(2)},q^{(2)}|\lambda)]
\end{equation}

Where the expectation, here, is no longer conditional on ($O,\lambda')$
but is now conditional on ($O^{(1)},O^{(2)},\lambda')$. If the observation
sequences are independent, this can be written as:

\begin{equation}
\mathcal{Q}(\lambda,\lambda')=E_{q^{(1)}|O^{(1)},\lambda'}[\log p(O^{(1)},q^{(1)}|\lambda)]+E_{q^{(2)}|O^{(2)},\lambda'}[\log p(O^{(2)},q^{(2)}|\lambda)]\label{eq:auxmult}
\end{equation}

We show how the EM algorithm can be simply adapted for this case by
considering the optimization of $a$. Writing out the terms dependent
on $a$, we obtain:

\begin{equation}
\sum_{t=2}^{T^{(1)}}E_{(q_{t-1}^{(1)},q_{t}^{(1)})}\left[\log p(q_{t}^{(1)}|q_{t-1}^{(1)},a)\right]+\sum_{t=2}^{T^{(2)}}E_{(q_{t-1}^{(2)},q_{t}^{(2)})}\left[\log p(q_{t}^{(2)}|q_{t-1}^{(2)},a)\right]
\end{equation}

Or just the following sum:

\begin{equation}
\sum_{k=1}^{2}\sum_{t=2}^{T^{(k)}}E_{(q_{t-1}^{(k)},q_{t}^{(k)})}\left[\log p(q_{t}^{(k)}|q_{t-1}^{(k)},a)\right]
\end{equation}

Through the same reasoning as employed in the previous section for
the sum in the 2nd term, the way to maximize this is simply to set:

\begin{equation}
a_{ij}=\frac{1}{\zeta_{i}}\sum_{k=1}^{2}\sum_{t=2}^{T^{(k)}}p(q_{t}^{(k)}=j,q_{t-1}^{(k)}=i|O^{(k)},\lambda')\label{eq:amult}
\end{equation}

Where $\zeta_{i}=\sum_{j=1}^{N}a_{ij}$, as before. A similar reasoning
is applied for $\pi$, which becomes proportional to $\sum_{k=1}^{2}P(q_{1}^{k)}=i|O^{(k)},\lambda')$.
The case for more than 2 independent observations is similar. Note
that \ref{eq:amult} is the same as the formula derived in \cite{Li2000b}.

\subsection*{Discussion}

In this article, we have presented an information-theoretic interpretation
of the EM algorithm for hidden Markov models and demonstrated its
reduction to simpler, known problems in statistical optimization.
We have also demonstrated the use of this method for training HMMs
on multiple observation sequences (a situation which arises widely
in practice). It is our hope that this new derivation method will
be used to provide more fundamental insights into the use of the EM
algorithm for various HMM-like models, and perhaps help lead intuition
to discovering efficient algorithms for training new custom HMM-like
models for various problems.

\bibliographystyle{plain}
\bibliography{visual,paper,gml}

\end{document}